\documentclass[%
reprint,
aip,
amsmath,amssymb,
jap
]{revtex4-1}

\usepackage{graphicx}
\usepackage{dcolumn}
\usepackage{bm}
\usepackage{textcomp} 
\usepackage[anythingbreaks]{breakurl}

\begin{document}

\title{Hybrid optical pumping of K and Rb atoms in a paraffin coated vapor cell}

\author{Wenhao~Li}
 \affiliation{State Key Laboratory of Advanced Optical Communication Systems and Networks, School of Electronics Engineering and Computer Science, and Center for Quantum Information Technology, Peking University, Beijing 100871, China.}

\author{Xiang~Peng}%
 \email[]{xiangpeng@pku.edu.cn}
 \affiliation{State Key Laboratory of Advanced Optical Communication Systems and Networks, School of Electronics Engineering and Computer Science, and Center for Quantum Information Technology, Peking University, Beijing 100871, China.}

\author{Dmitry~Budker}%
 \affiliation{Helmholtz Institute Mainz, 55099 Mainz, Germany.}
 \affiliation{Department of Physics, University of California, Berkeley, California 94720-7300, USA.}
 \affiliation{Nuclear Science Division, Lawrence Berkeley National Laboratory, Berkeley, California 94720, USA.}

\author{Arne~Wickenbrock}%
 \affiliation{Helmholtz Institute Mainz, 55099 Mainz, Germany.}
 \affiliation{Johannes Gutenberg-University Mainz, 55128 Mainz, Germany}
 
\author{Bo~Pang}%
 \affiliation{State Key Laboratory of Advanced Optical Communication Systems and Networks, School of Electronics Engineering and Computer Science, and Center for Quantum Information Technology, Peking University, Beijing 100871, China.}

\author{Rui~Zhang}%
 \affiliation{State Key Laboratory of Advanced Optical Communication Systems and Networks, School of Electronics Engineering and Computer Science, and Center for Quantum Information Technology, Peking University, Beijing 100871, China.}
 \affiliation{College of Science, and Interdisciplinary Center for Quantum Information, National University of Defense Technology, Changsha 410073, China.}
 
\author{Hong~Guo}%
 \email[]{hongguo@pku.edu.cn}
 \affiliation{State Key Laboratory of Advanced Optical Communication Systems and Networks, School of Electronics Engineering and Computer Science, and Center for Quantum Information Technology, Peking University, Beijing 100871, China.}

\date{\today}

\begin{abstract}
Dynamic hybrid optical pumping effects with a radio-frequency-field-driven nonlinear magneto-optical rotation (RF NMOR) scheme are studied in a dual-species paraffin coated vapor cell. By pumping K atoms and probing $^{87}$Rb atoms, we achieve an intrinsic magnetic resonance linewidth of 3~Hz and the observed resonance is immune to power broadening and light-shift effects. Such operation scheme shows favorable prospects for atomic magnetometry applications.
\end{abstract}

\maketitle


Creation of atomic polarization with optical pumping \cite{Happer1972} is one of the most important techniques in atomic magnetometry. Due to the conservation of angular momentum (AM), the process of optical pumping transfers AM from photons to atoms through light-atom interactions and creates detectable macroscopic polarization in an atomic ensemble under various experimental configurations. In addition to the transfer of AM from photons to atoms, AM can also be transferred between atoms via spin-exchange collisions (SECs) \cite{Dehmelt1958}, where two colliding atoms exchange their electron spins while the total spin is conserved. Such process is often referred to as spin-exchange optical pump (SEOP) \cite{Walker1997}. SEOP can be used not only to polarize noble-gas nuclei \cite{Babcock2003} and metastable helium atoms \cite{Blinov1979} but also to achieve polarization transfer between different alkali species, where this technique is also called hybrid optical pumping \cite{Romalis2010}.

The sensitivity of an atomic magnetometer benefits from a narrow magnetic-resonance linewidth (or equivalently, long Zeeman-coherence lifetime) and a high signal-to-noise ratio (SNR). The resonance linewidth is limited by several decoherence mechanisms, such as SECs, wall collisions, magnetic-field gradients, etc. There are two commonly used ways to preserve the coherence lifetime of atomic polarization. One is to use a buffer gas to slow down atomic diffusion to reduce collisions with the cell wall, the other is to coat the inner surface of the cell with anti-relaxation wall coating (AWC) which allows thousands to a million \cite{Balabas2010PRL} of bounces without depolarization. For magnetometric applications, AWC cells have several advantages over buffer gas cells, such as the ability of operating at lower light power, reduced sensitivity to magnetic-field gradients \cite{Pustelny2006} and narrower spectral linewidth \cite{Balabas2010}. Operating at high temperatures is one way to achieve a larger signal amplitude and thus a better SNR. Although most AWC, especially paraffin wax, have limited operating temperatures due to low melting point of the coating material ($\sim 60^{\circ}$C), recent study shows that, with careful temperature control, paraffin-coated vapor cells can operate at temperatures above the coating material's melting point and exhibit an improved fundamental sensitivity limit \cite{Li2016}. The other way to improve SNR is to create larger atomic polarization by using higher light power but then one has to contend with possible power-broadening and light-shift effects degrading the sensitivity. It is reported previously that the power broadening and light-shift effects can be mitigated by optically pumping and probing different ground-state hyperfine manifolds so that the probed levels are barely affected by the pump light \cite{Chalupczak2010,Zhivun2014}.

In this letter, we demonstrate hybrid optical pumping of $^{87}$Rb atoms and investigate magnetic resonances in an radio-frequency-field-driven nonlinear magneto-optical rotation (RF NMOR) scheme \cite{Zigdon2010} in a home-made paraffin-coated K-$^{87}$Rb vapor cell. By comparing the direct-pumping scheme, i.e. optically pumping and probing K, with hybrid optical pumping in terms of their responses to various experimental parameters (light power, light detuning, temperature), we show that hybrid-optical-pumping-induced magnetic resonance exhibits no measurable power-broadening or light-shift effects, which is distinct from previous techniques \cite{Chalupczak2010}. The narrowest resonance linewidth for hybrid optical pumping scheme in our dual-species vapor cell is 3~Hz, which is comparable to the best single-species paraffin-coated vapor cells reported previously \cite{Balabas2010,Li2016,Balabas1999}. Therefore, such operation scheme shows favorable prospects for the future high-sensitivity magnetometer applications.

\begin{figure}[thbp]
\centering
\fbox{\includegraphics[width=\linewidth]{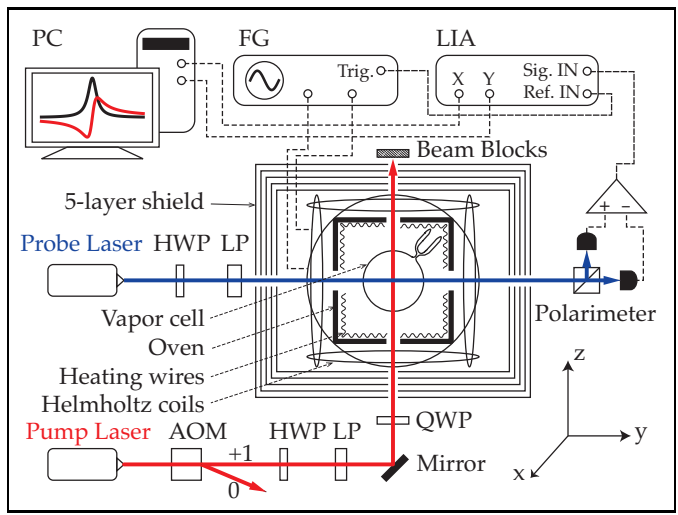}}
\caption{Schematic of the experimental setup. PC: personal computer, FG: function generator, LIA: lock-in amplifier, HWP: half-wave plate, QWP: quarter-wave plate, LP: linear polarizer, AOM: acousto-optic modulator.}
\label{fig:setup}
\end{figure}

The heart of the experimental setup is a paraffin-coated vapor cell which contains K in natural abundance and isotopically enriched $^{87}$Rb. The cell is spherical in shape with a diameter of 45~mm. Two separate 30~mm-long stems, which contain K and $^{87}$Rb respectively, connect to the cell body via a single opening. In principle, the double-stem design allows independent temperature control of the two atomic species so that it is possible to control the ratios of vapor density for the two alkali metals [see Fig.~\ref{fig:setup}]. Although in the long term atomic diffusion leads to mixing of the two reservoirs, the specific timescale of such transition is estimated to be on the order of several months to years, which is much longer than the measurement time. The capillaries that connect the cell body and the stems are kept thin enough to reduce relaxation from atom exchange between the atomic vapor and the reservoir. The coating material is commercial paraffin wax with a melting temperature of about 55$^{\circ}$C. Detailed characterization of vapor cells with the similar coating material can be found in \cite{Li2016}.

Figure \ref{fig:setup} shows the schematic of the experiment setup. The vapor cell is contained in a Teflon oven that can be heated with quad-twisted copper wires carrying 100~kHz AC current. The AC current creates oscillating magnetic fields far beyond the bandwidth of the magnetometer with time average zero and therefore does not perturb the magnetic field measurement. A thermocouple is placed inside the oven $\sim 1.5$~cm away from the vapor cell to measure the temperature of the air which is considered to be close to the temperature of the cell surface. A five-layer $\mu$-metal shield provides $10^6$ shielding factor for the environmental magnetic fields. Three sets of coils inside the shield generate magnetic fields needed for the RF NMOR experiment. The $\hat{z}$-coil is driven by a DC current source and generates the leading magnetic fields $B_0$ in the experiment. The $\hat{x}$- and $\hat{y}$-coils are driven by two synchronized sinusoidal signals with adjustable phase shift which is generated by a function generator. By adjusting the relative phase shift between the two signals, we can create a rotating or counter-rotating magnetic field $B_1(\cos{\omega_{\mathrm{rf}} t} \pm i\sin{\omega_{\mathrm{rf}} t})$ in the $\hat{x}$-$\hat{y}$ plane. During the experiment, $B_1$ is always kept low enough ($\sim 0.1$~nT) to avoid the influence of RF broadening. Pump light is generated with a K D1 (770~nm) distributed feedback (DFB) diode laser whose power can be changed by adjusting the modulation level of the AOM (acousto-optic modulator). The circularly polarized pump light propagates along $\hat{z}$ and creates atomic polarization in the direction of the leading field $B_0$. The source of probe light can be switched between a K D2 (766~nm) DFB diode laser and a Rb D2 (780~nm) external cavity diode laser (ECDL). Linearly polarized probe light propagates along $\hat{y}$ and the rotation of its polarization axis is measured with a polarimeter. Both the trigger signal from the function generator (used as reference) and the measured optical rotation signal are fed into a lock-in amplifier (LIA) to extract the in-phase and the quadrature component of the magnetic-resonance signal.

\begin{figure}[t!]
\centering
\fbox{\includegraphics[width=\linewidth]{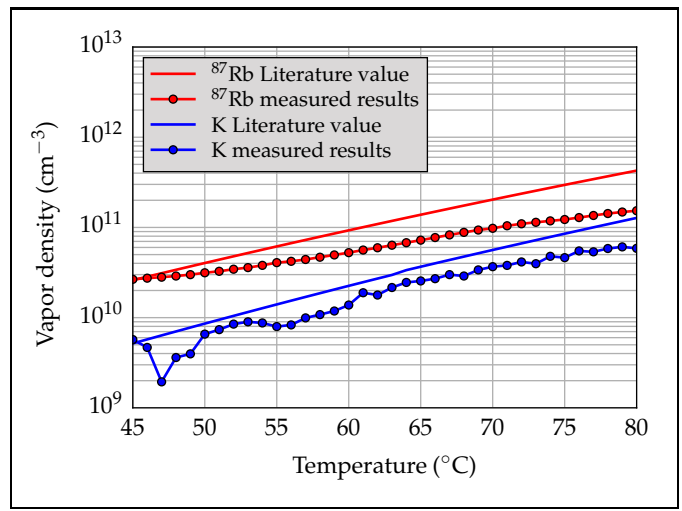}}
\caption{Results of vapor density measurements. The absorption profile is taken and then fitted to extract the vapor density \cite{ADMpackage}. The power of the probe beam is kept at $\sim 4$~\textmu W to avoid nonlinear effects due to probe-light pumping effect.}
\label{fig:vp}
\end{figure}

\begin{figure*}[t!]
  \centering
  \fbox{\includegraphics[]{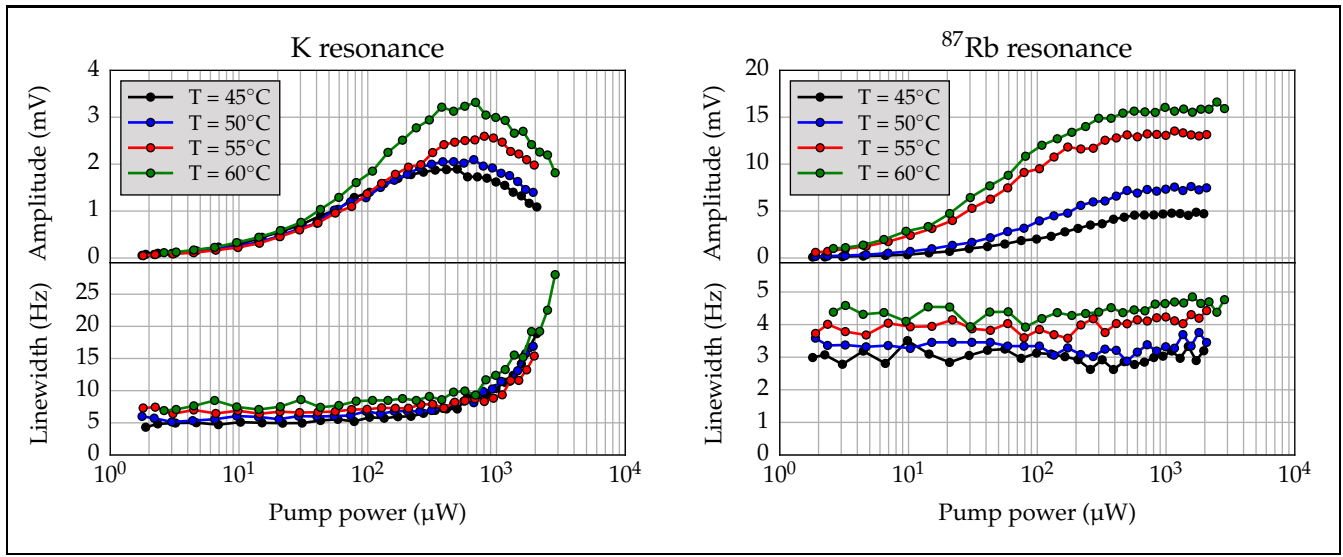}}
  \caption{Pump power dependence of magnetic resonance for $^{87}$Rb and K. During the measurements, the pump light is always tuned near the center of K D1 resonance to maximize the resonance amplitude and its power is altered by adjusting the modulation voltage level of AOM. The frequency of the probe light is also tuned to the point where resonance amplitude is maximized.}
  \label{fig:pd}
\end{figure*}

In order to measure the atomic vapor density of both K and $^{87}$Rb in the vapor cell, we measure absorption by summing the two output channels from the balanced detector while scanning the laser detuning across $^{87}$Rb (or K) D2 transitions. Figure~\ref{fig:vp} shows the results of the vapor density measurements. It should be noted that, during the experiment, there is no separate temperature control for the two stems so the temperature that determines the vapor density in the cell body should be nominally the same for K and $^{87}$Rb. The fitting error for the K vapor density at low temperatures is relatively large due to the limited SNR. The measured results for both atomic species are about 20\% -- 80\% of their saturated vapor density, which may be due to the absorption of alkali atoms by the coating material \cite{Balabas2012,budker2013optical}. The discrepancy between the measured value and the literature value varies with temperature. The reason for this could be the change of the temperature difference between the cell body and the stem during the continuous heating process.

One way to induce spin precession in an atomic ensemble is to use a continuous pump beam to create atomic polarization along the leading-field direction and apply a small rotating magnetic field at the Larmor frequency in the transverse plane to drive the resonance \cite{Groeger2006,Schultze2012,Chalupczak2010}. Another way is to modulate the amplitude, frequency or polarization of a transverse pump beam at the Larmor frequency (or harmonics of the Larmor frequency depending on the specific experimental configuration) so that the pump beam synchronously drives the precessing polarization in the transverse plane \cite{Bell1961,Gawlik2006,Budker2002a,Grujic2013}. Both methods work well for single-species experiments. But for a dual-species experiment such as reported in this letter, synchronous optical pumping cannot achieve a high efficiency of AM transfer between $^{87}$Rb and K via SECs at relatively large magnetic fields ($>1000~\mathrm{nT}$) due to the different nuclear $g$-factors of the two species. In the case of $^{39}$K and $^{87}$Rb, the difference in their precession frequencies is estimated to be on the order of 10~Hz at a magnetic field of 1000~nT, which is larger than the intrinsic magnetic resonance linewidth for both atoms [see Fig.~\ref{fig:pd}]. As a consequence, the precession of one atomic species will fall behind the other and eventually such frequency mismatch will limit the maximum polarization that is transferred via SECs.

For this reason, we use a continuous pump beam to polarize the K atoms along $B_0$ and $^{87}$Rb is polarized along the same direction via SECs with K. The magnetic resonance is realized by applying a rotating RF field $B_1(\omega_{\mathrm{rf}})$ that tilts both the $^{87}$Rb and K magnetization away from the leading-field direction and causes the transverse component of the magnetization to precess around $B_0$ at the Larmor frequency. Figure~\ref{fig:pd} shows the pump power dependence of the resonance amplitude and linewidth for both K and $^{87}$Rb. We can see from the figure that there is a saturation point for the pump power. When the pump power is below the saturation point, K atoms receive AM from the pump photons and the longitudinal magnetization of K increases with pump power. Meanwhile, due to fast SECs, atomic polarization is continuously transferred from K to $^{87}$Rb so that the amplitude of $^{87}$Rb resonance also increases with pump power. However, when the pump power is beyond the saturation point, extra pump photons cannot create any more longitudinal atomic polarization for K. On the contrary, they interact with K atoms that are precessing around $B_0$ and thus destroy the transverse polarization of K atoms. Hence, in this regime, the amplitude of K resonance decreases with pump power. But for $^{87}$Rb, since it does not directly interact with the pump light, extra pump photons cannot disturb its transverse polarization. The amplitude of $^{87}$Rb resonance becomes flat when the pump power exceeds the saturation point. As for the resonance linewidth, below the saturation point, both K and $^{87}$Rb resonances show no dependence on the pump power. However, above the saturation point, K resonance is disturbed by the extra pump light photons and relaxation due to the pump light becomes the limiting factor for the lifetime of K Zeeman coherence. For the linewidth of $^{87}$Rb resonance, it stays almost flat within the pump power range explored in the experiment and shows no obvious power broadening even beyond the saturation point.

\begin{figure}[t!]
\centering
\fbox{\includegraphics[width=\linewidth]{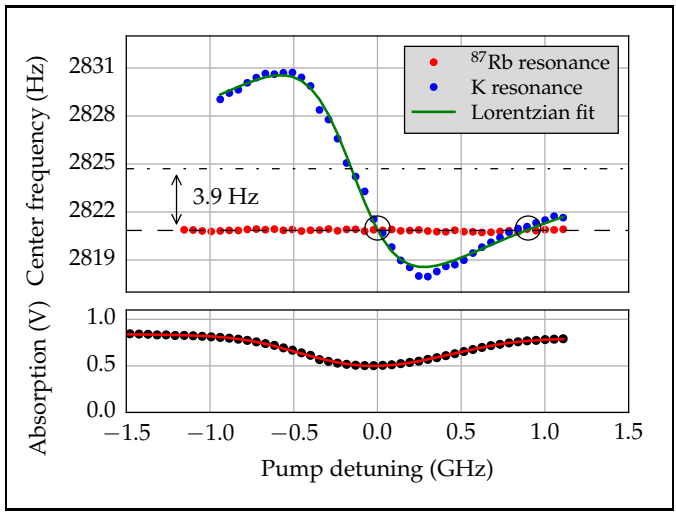}}
\caption{Pump light detuning dependence of resonance frequency for K and $^{87}$Rb (top) and K absorption profile (black dot) together with linear absorption fitting (red line) (bottom). The power of the pump light was 100~\textmu W and the data were collected at $45^{\circ}$C. The dashdot line marks the zero-light-shift center frequency of the K resonance. The dashed line marks the mean value of the measured $^{87}$Rb center frequencies. Two circles in the upper plot mark the ``magic'' detuning points where $^{87}$Rb and K precess at the same frequency.}
\label{fig:pump}
\end{figure}

In order to study the temperature dependence of the magnetic resonances for K and $^{87}$Rb, the temperature of the cell is varied from $45^{\circ}$C to $60^{\circ}$C during the measurement. The highest temperature is chosen to be high enough to yield a significant signal amplitude and low enough to avoid degradation of the coating material. We can observe from Fig.~\ref{fig:pd} that the resonance amplitude of both K and $^{87}$Rb increase with temperature, which can be explained by the increase of atomic vapor density. Regarding the linewidth, the spin-exchange broadened linewidth can be estimated by the equation $2\pi\cdot\Delta\nu_{\mathrm{ex}}=\gamma_{\mathrm{ex}}=n\sigma_{\mathrm{ex}}v/q$, where $n$ is the atomic vapor density, $\sigma_{\mathrm{ex}}$ is the spin-exchange cross section, $v$ is the mean relative thermal velocity, and $q$ is the appropriate nuclear slowing-down factor \cite{budker2013optical}. In our experiment, the spin-exchange broadened linewidth is $\Delta\nu_{\mathrm{ex}} \approx 1~\mathrm{Hz}$. Therefore, SECs are not the dominant factor for the resonance linewidth in the explored temperature range. Increasing the temperature enhances SECs and changes the quality of the coating material, which leads to further broadening of the linewidth. We also observe that, compared with direct pumping and probing of K, the $^{87}$Rb resonance with hybrid optical pumping not only has the advantage of being free from power broadening but also has a narrower observed intrinsic resonance linewidth (3 Hz) than the K resonance. At high pump powers, both the resonance amplitude and linewidth of the $^{87}$Rb magnetic resonance have a flat response to K pump power variations. This is beneficial to magnetometry applications where both insensitivity to power-fluctuations and narrow linewidth are required.

Figure~\ref{fig:pump} shows the dependence of both magnetic resonance center frequencies on the detuning of the K pump laser. Due to the vector light shift \cite{Mathur1968} induced by the circularly polarized pump light along $B_0$, the K resonance exhibits a Lorentzian-shaped dependence on the pump detuning. The dispersive Lorentzian fit has an amplitude of 12.87~Hz and a width of 0.91~GHz. However, for the $^{87}$Rb resonance, the center frequency shows essentially no dependence on the pump power which can be explained by the fact that there is no significant interaction between $^{87}$Rb atoms and the K D1 pump light. It should be noted that the difference between the mean center frequency of $^{87}$Rb and the center frequency of K at zero-light-shift point is about 3.9~Hz, which results from the different nuclear $g$-factors between $^{39}$K and $^{87}$Rb. Therefore, another benefit from pumping K and probing $^{87}$Rb is the immunity to light shift effect. We also find that the precession frequency of $^{87}$Rb and K match with each other at two ``magic'' pump laser detuning frequencies. Since K resonance is insensitive to frequency fluctuations of the pump light at the ``turning point'' of the detuning dependence curve, by properly adjusting the power and detuning of the pump laser, we are able to shift the ``turning point'' to the ``magic'' frequency point so that the K and $^{87}$Rb resonances share the same Larmor frequency and both of them are insensitive to frequency fluctuations of the pump laser.

In conclusion, we demonstrate hybrid optical pumping of $^{87}$Rb atoms in a paraffin-coated K--$^{87}$Rb vapor cell. The measured RF NMOR resonances show that by optically pumping K atoms and probing $^{87}$Rb atoms, we can achieve larger resonance amplitudes compared with probing K atoms and narrower resonance linewidths (3~Hz) which are free from power broadening. We also show that $^{87}$Rb resonances are immune to light shifts originated from the pump beam. Such operation scheme allows for the use of high-power pump laser without the sacrifice of additional power broadening and light shift effects, which is of essential importance to high-sensitivity magnetometry applications. The results in this letter should be, in principle, valid for other alkali metal combinations. Our future work includes applying separate temperature control for the two cell stems and testing the performance of the dual-species paraffin-coated vapor cell in a practical magnetometer system.

\begin{acknowledgments}
The authors gratefully acknowledge M.V. Balabas, J. Stalnaker, and E.~Zhivun for helpful discussions and thank S. Li for his contribution to the construction of the early-stage experimental setup. X.P. and H.G. acknowledge support by the National Natural Science Foundation of China (61571018, 61531003). H.G. acknowledges support by the National Science Fund for Distinguished Young Scholars of China (61225003). D.B. acknowledges support by the QUTEGA program.
\end{acknowledgments}

\bibliography{sample.bib}

\end{document}